\documentclass{webofc}
\usepackage[varg]{txfonts}   
\usepackage{xcolor}

\newcommand{\eq}[1]{\begin{align} #1 \end{align}}

\begin{document}
\title{Spinodal enhancement of fluctuations in nucleus-nucleus collisions}

\author{\firstname{Roman} \lastname{Poberezhnyuk}\inst{1,2}\fnsep\thanks{\email{rpoberezhnyuk@bitp.kiev.ua, poberezhnyuk@fias.uni-frankfurt.de}} \and
        \firstname{Oleh} \lastname{Savchuk}\inst{3,1,2,4} \and
        \firstname{Volodymyr} \lastname{Vovchenko}\inst{5} \and
        \firstname{Volodymyr} \lastname{Kuznietsov}\inst{1,5,2} \and
        \firstname{Jan} \lastname{Steinheimer}\inst{2} \and
        \firstname{Mark} \lastname{Gorenstein}\inst{1,2} \and
        \firstname{Horst} \lastname{Stoecker}\inst{2,4,6}
}

\institute{Bogolyubov Institute for Theoretical Physics, 03680 Kyiv, Ukraine
\and
           Frankfurt Institute for Advanced Studies, Giersch Science Center, Ruth-Moufang-Str. 1, D-60438 Frankfurt am Main, Germany 
\and
           Facility for Rare Isotope Beams, Michigan State University, East Lansing, MI 48824 USA
\and
           GSI Helmholtzzentrum f\"ur Schwerionenforschung GmbH, Planckstr. 1, D-64291 Darmstadt, Germany
\and
           Physics Department, University of Houston, Box 351550, Houston, TX 77204, USA
\and
           Institute of Theoretical Physics, Goethe Universit\"at, Frankfurt, Germany
          }

\abstract{%
  Subensemble Acceptance Method (SAM)~\cite{Vovchenko:2020tsr,Poberezhnyuk:2020ayn} is an essential link between measured event-by-event fluctuations and their grand canonical theoretical predictions such as lattice QCD. The method allows quantifying the global conservation law effects in fluctuations. In its basic formulation, SAM requires a sufficiently large system such as created in central nucleus-nucleus collisions and sufficient space-momentum correlations. Directly in the spinodal region of the First Order Phase Transition (FOPT) different approximations should be used that account for finite size effects. Thus, we present the generalization of SAM applicable in both the pure phases, metastable and unstable regions of the phase diagram~\cite{Kuznietsov:2023iyu}. Obtained analytic formulas indicate the enhancement of fluctuations due to crossing the spinodal region of FOPT and are tested using molecular dynamics simulations. A rather good agreement is observed. Using transport model calculations with interaction potential we show that the spinodal enhancement of fluctuations survives till the later stages of collision via the memory effect~\cite{Savchuk:2022msa}. However, at low collision energies the space-momentum correlation is not strong enough for this signal to be transferred to second and third order cumulants measured in momentum subspace. This result agrees well with recent HADES data on proton number fluctuations at $\sqrt{s_{NN}}=2.4$
 GeV which are found to be consistent with the binomial momentum space acceptance~\cite{Savchuk:2022ljy}.
}
\maketitle
\section{Introduction}
\label{intro}

In heavy ion collisions (HIC) and their microscopic (e.g., transport model) simulations fluctuations are measured within a subsystem which is comparable in size to the size of the total system.
This situation is described by the new statistical ensemble, called subensemble~\cite{Vovchenko:2020tsr,Poberezhnyuk:2020ayn}.
On the other hand, in theoretical calculations (for instance Lattice QCD and statistical models) fluctuations are calculated within grand canonical ensemble (GCE) which is a limiting case of the subensemble when the total system is much larger than the observed subsystem, and thus the effects of global charge conservation are negligible.
While ensembles are equivalent w.r.t. mean quantities they are not equivalent w.r.t. fluctuations \cite{Begun:2004pk}, thus they cannot be directly compared to each other. The Subensemble Acceptance Method (SAM) \cite{Vovchenko:2020tsr} provides the
model-independent analytic connection
between fluctuations in subensemble and fluctuations calculated within GCE.
The method is applicable for sufficiently large systems such as those created in central HIC.
As an example, we present the SAM formulas for intensive fluctuation measures of conserved charge distribution in the subsystem up to 4-th order in the case of a single conserved charge:
\eq{\label{eq:w}
\omega  = \beta \, \omega_{GC}~,~~~~~~~~~ 
S\sigma  = (1-2\alpha) \, S\sigma_{GC}~,~~~~~~~~~ 
\kappa\sigma^2  = (1-3\alpha \beta) \, \kappa\sigma^2_{GC} - 3 \alpha \beta \left( S\sigma^2_{GC}\right)^2.
}
Here $\omega$, $S\sigma$, and $\kappa\sigma^2$ are the scaled variance, skewness, and kurtosis of a conserved charge distribution, $\omega_{GC}$, $S\sigma_{GC}$, and $\kappa\sigma_{GC}^2$ are the corresponding GCE values,  $\alpha$ is the acceptance parameter equal to the fraction of the total volume occupied by the subvolume, $\beta\equiv 1-\alpha$. The GCE and canonical ensemble limits correspond to $\alpha\rightarrow 0$ and $\alpha\rightarrow 1$, respectively. For the case of a noninteracting gas, the SAM formulas are reduced to binomial acceptance formulas~\cite{Bzdak:2012ab,Savchuk:2019xfg}.

Various generalizations and extensions of SAM include extensions to arbitrary number of conserved charges \cite{Vovchenko:2020gne}, factorial cumulants \cite{Barej:2022jij}, nonuniform systems \cite{Vovchenko:2021yen},
the next-to-leading order corrections in the presence of finite size effects have been obtained \cite{Barej:2022ccb}.
The SAM formulas have been checked numerically for the specific examples of a system containing FOPT and the critical point (CP) --- the well-known Van der Waals mean-field model~\cite{Poberezhnyuk:2020ayn} and molecular dynamics simulations of the classical 
system of particles with
Lenard-Jones (LJ) interaction potential in a box~\cite{Kuznietsov:2022pcn}. 
In these studies only pure phase and supercritical regions of the phase diagram were considered.
Here we focus on the application of the method at low collision energies and, in particular, in the mixed phase of FOPT.

\section{Fluctuations in the mixed phase of first order phase transition}
\label{sec-2}

\begin{figure}[h!]
\centering
\sidecaption
    \includegraphics[width=.55\textwidth]{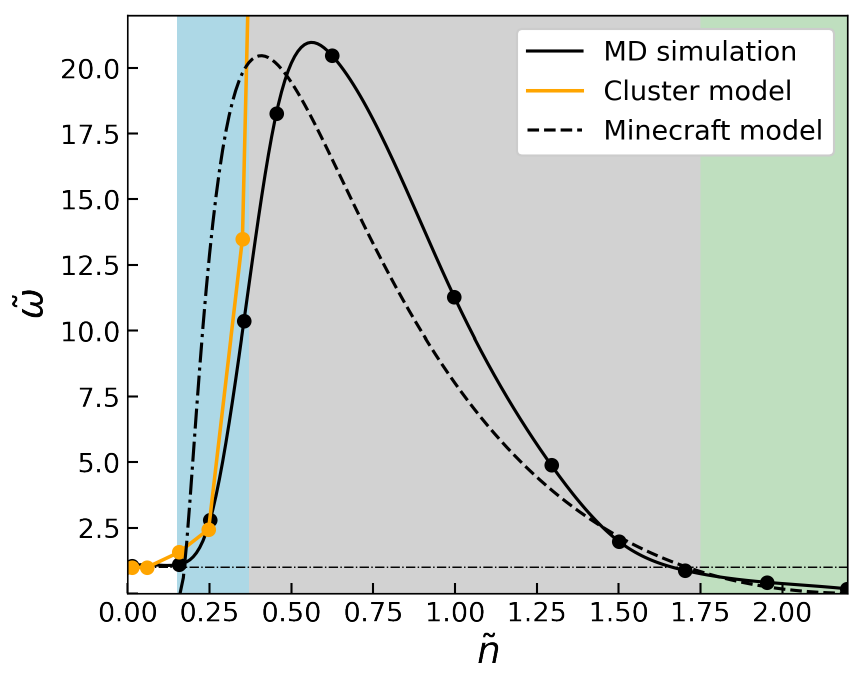}
    \caption{
    Scaled variance of particle number distribution inside subvolume for $\alpha =0.2$ in the mixed phase ($T/T_c=0.76$). The solid line corresponds to the molecular dynamics result. The orange line shows the cluster model results in the nucleation region  $0.16\le n/n_c\le 0.35$. The dashed line shows the results of the Minecraft model in the spinodal region $0.35\le  n/n_c\le 1.75$, and the dash-dotted line is its 
    extension to the nucleation region. 
    }
    \label{fig:w00}
\end{figure}

For a description of the mixed phase we used~\cite{Kuznietsov:2023iyu} molecular dynamics simulations with
LJ potential, similar to Ref.~\cite{Kuznietsov:2022pcn}.
Figure \ref{fig:w00} shows the scaled variance $\tilde{\omega}=\omega/(1-\alpha)$ of particle number distribution in the central cell with $\alpha=0.2$ corrected for particle number conservation using SAM.  

We find that in the region of the phase diagram which corresponds to the first metastable state, scaled variance is well described by the model of noninteracting clusters in GCE (with the distribution of clusters taken from the simulation itself). Thus, the SAM is applicable. 

In the spinodal region, the particle number distribution exhibits double peaks and corresponding large fluctuations. These fluctuations arise primarily from geometric factors. To investigate this phenomenon, a geometric model was developed, representing a cubic liquid mass moving within a cubic volume.
Termed the Minecraft model, this approach simplifies complex forms into cubic geometries. Subsequent analysis of various shapes indicates that, in terms of scaled variance, this simplification provides a reasonable approximation. 
Assuming a uniform static system in the mixed phase, the formalism is reduced to the model-independent formulas for cumulants obtained earlier~\cite{Poberezhnyuk:2020cen}. 
However, our analysis indicates that the uniform approximation cannot be used for realistic systems such as those created in HIC.

One can see that two simple analytical approximations, cluster model and Minecraft model, can describe the numeric result rather well.

\section{Fluctuations in heavy ion collisions}
\label{sec-2}
Contrary to the scenario discussed in Sec.~\ref{sec-2} in HIC the created system is expanding and fluctuations are measured not in a coordinate but in a momentum subspace.
Thus, for a more realistic description we used~\cite{Savchuk:2022msa} the UrQMD transport model with interaction potential~\cite{OmanaKuttan:2022the}\footnote{Namely, the Chiral SU(3)-flavor parity-doublet Polyakovloop quark-hadron mean-field model (CMF) potential~\cite{Steinheimer:2010ib,Steinheimer:2011ea} in its most recent version~\cite{Motornenko:2019arp}}
and studied proton number fluctuations in the central cell of volume 27~fm$^3$. We artificially introduced the FOPT by modifying the potential to create a minimum in energy density (see~\cite{Savchuk:2022msa} for details). The two potentials (with and without PT) are indistinguishable at densities less than 2.5 nuclear saturation density and system trajectories are similar~\cite{Savchuk:2022msa}. 
In the case of the PT the phase separation is observed with two maxima of particle number distribution while crossing the spinodal region. The corresponding strong enhancement in fluctuations survives until the later stages of collision, where two potentials are indistinguishable. 

However, in contrast to a coordinate subspace, in a momentum subspace we do not observe the strong signal of the PT in scaled variance, $\omega[p]$, and skewness, $S\sigma[p]$, of proton number distribution. This is because at low collision energies considered here, the collective flow is insufficient to produce strong correlations between the coordinate and momenta of particles and thus, the signal of spatial interactions between particles in fluctuations is not observed when measured in momentum subspace. Only in the higher moment, namely kurtosis, $\kappa\sigma^2[p]$, we begin to see the substantial difference in fluctuations between the two cases (with and without PT).

\begin{figure*}
\includegraphics[width=.31\textwidth]
{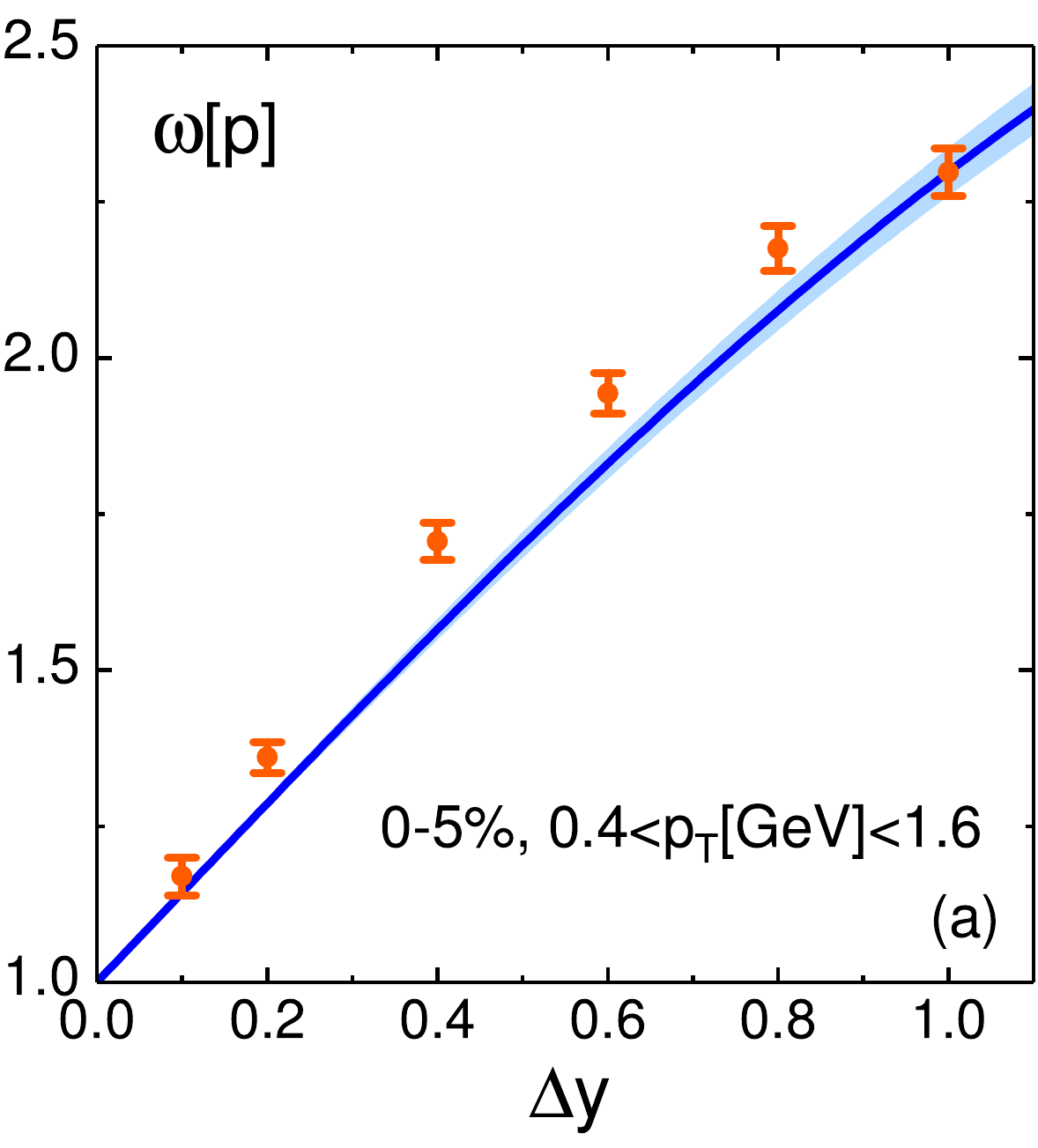}
\includegraphics[width=.32\textwidth]{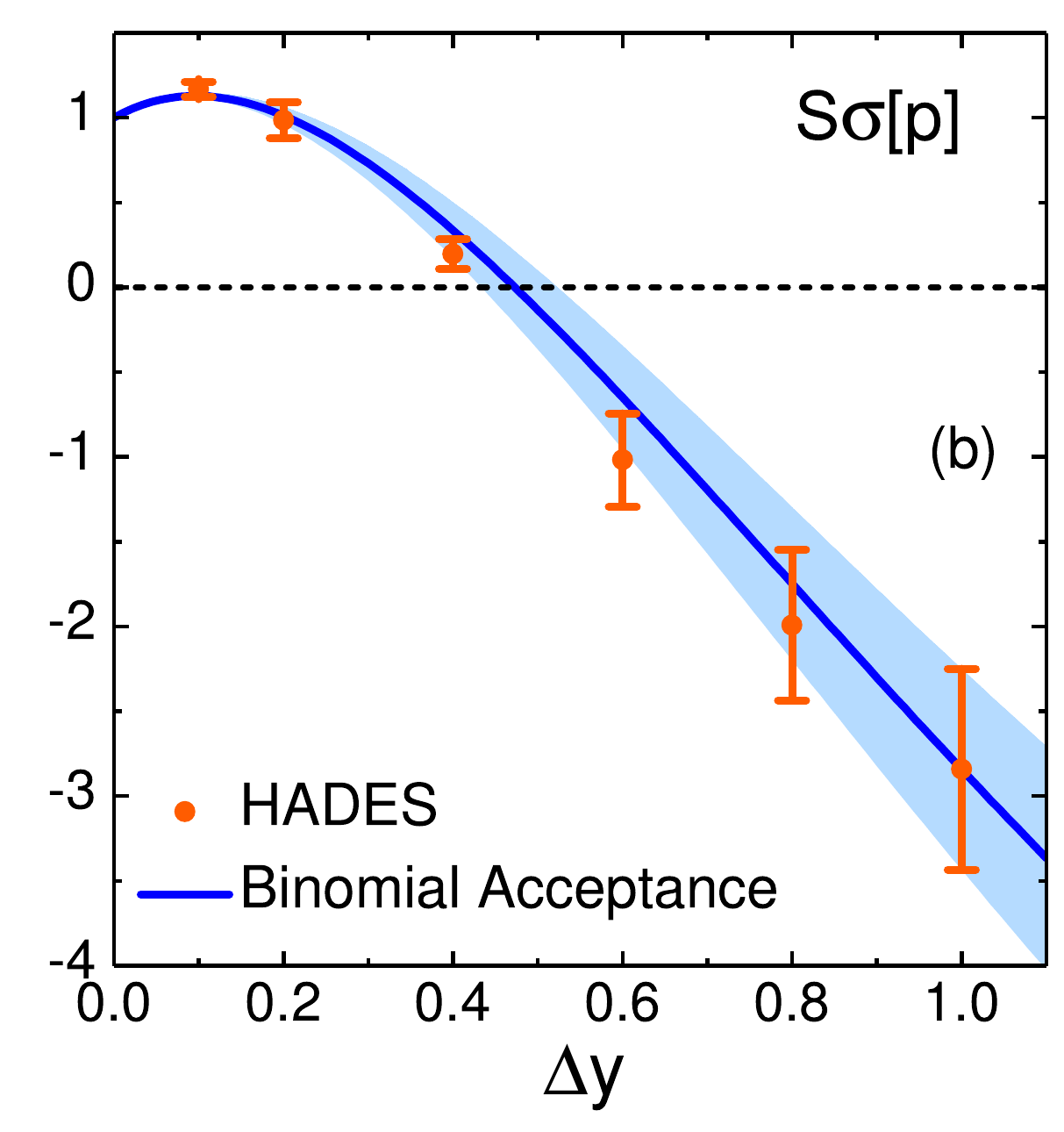}
\includegraphics[width=.32\textwidth]{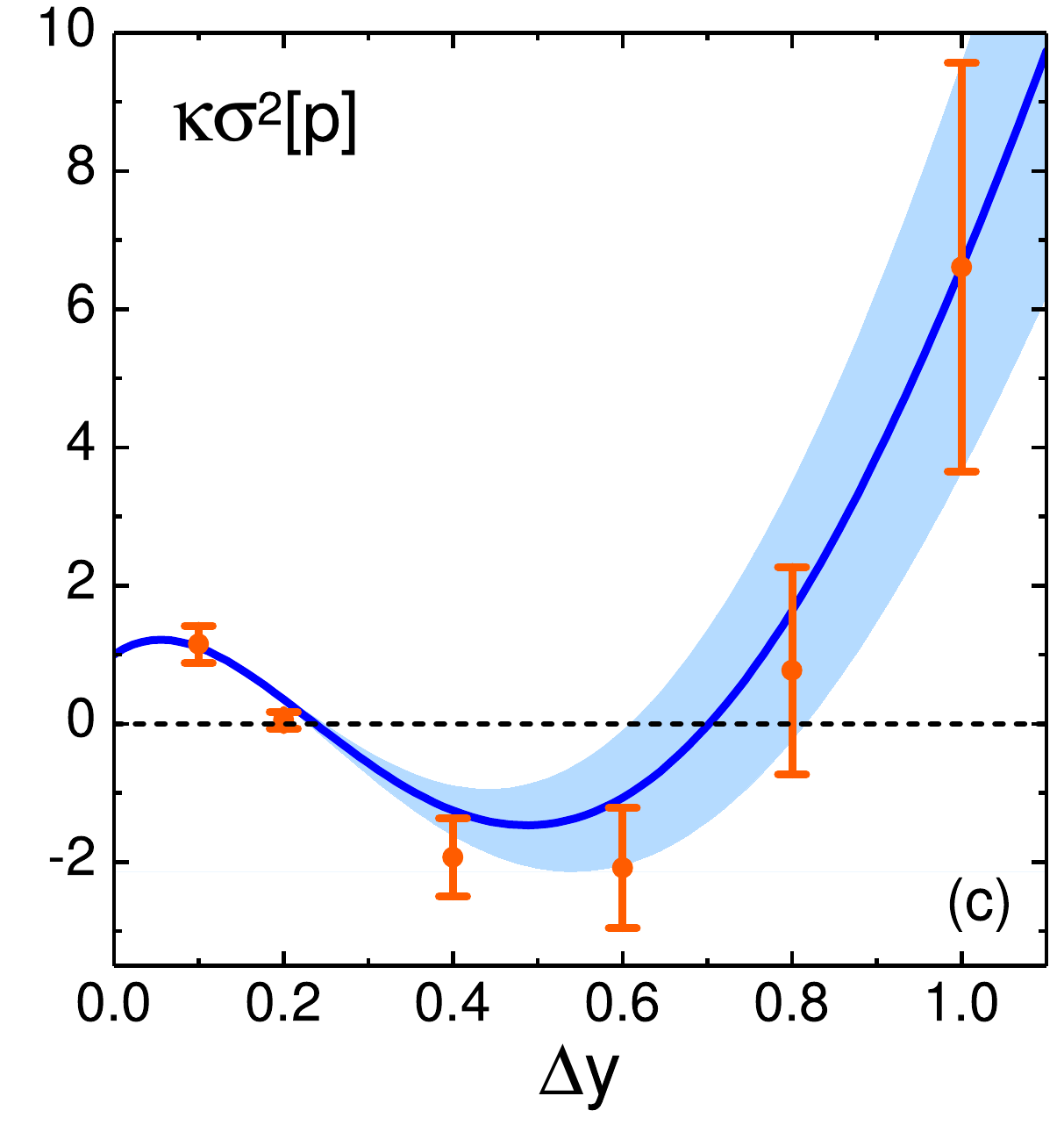}
\caption{Scaled variance (a), skewness (b), and kurtosis (c) of proton number distribution as functions of the rapidity interval $\Delta y$. The HADES data are shown by the symbols.  The line corresponds to the binomial acceptance formulas. The blue bands represent uncertainties due to HADES data errors in the $\Delta Y = 1$ rapidity interval.
}\label{fig:hades}
\end{figure*}

Considering this result, the question arises about the interpretation of the recent HADES data~\cite{HADES:2020wpc} which exhibit large proton number fluctuations at $\sqrt{s_{NN}}=2.4$~ GeV.
We show~\cite{Savchuk:2022ljy} that the $\Delta y$ dependencies of 
$\omega[p]$, $S\sigma[p]$, and $\kappa\sigma^2[p]$
obtained by HADES are well described by binomial acceptance formulas which assume that particle momenta are uncorrelated, see Fig.~\ref{fig:hades}. As an input, we use the experimental values of
$\omega[p]$, $S\sigma[p]$, and $\kappa\sigma^2[p]$
in the largest rapidity window of $\Delta Y = 1$ together with their uncertainties to calculate the corresponding values at $\Delta y < 1$ from binomial acceptance formulas.\footnote{We obtain the dependence of $\alpha$ on $\Delta y$ by fitting the HADES rapidity distribution by Gaussian function.}
This indicates that large fluctuations at $\Delta Y = 1$ are not local in rapidity space and their origin is to be determined.
For example, the event-by-event nucleon participant fluctuations can deviate from the estimate or
the presence of nuclear fragments can cause fluctuations in the number of bare protons, which can be quantitatively checked in future studies.

\section{Summary}
\label{sec-summary}

Subensemble acceptance method (SAM) formulas allow to compare fluctuations measured in different subensembles with each other and with grand canonical calculations.
 The method is applicable in the stable and metastable regions of a phase diagram. Directly in the vicinity of the critical point or in the spinodal region the modifications of the SAM which account for finite size effects should be used.
The expanding system created in heavy ion collisions exhibits large fluctuations when crossing the spinodal region. This signal survives until the later stages of a collision via the memory effect.
However, at low collision energies this signal is not transferred to second and third order cumulants measured in momentum subspace. 
This conclusion is in agreement with recent HADES data on proton number fluctuations at $\sqrt{s_{NN}}=2.4$~GeV which are consistent with the binomial limit of SAM.


\begin{thebibliography}{}


\bibitem{Vovchenko:2020tsr}
V.~Vovchenko, O.~Savchuk, R.~V.~Poberezhnyuk, M.~I.~Gorenstein and V.~Koch,
Phys. Lett. B \textbf{811}, 135868 (2020)

\bibitem{Poberezhnyuk:2020ayn}
R.~V.~Poberezhnyuk, O.~Savchuk, M.~I.~Gorenstein, V.~Vovchenko, K.~Taradiy, V.~V.~Begun, L.~Satarov, J.~Steinheimer and H.~Stoecker,
Phys. Rev. C \textbf{102}, no.2, 024908 (2020)

\bibitem{Kuznietsov:2023iyu}
V.~A.~Kuznietsov, O.~Savchuk, R.~V.~Poberezhnyuk, V.~Vovchenko, M.~I.~Gorenstein and H.~Stoecker,
Phys. Rev. C \textbf{107}, no.5, 055206 (2023)

\bibitem{Savchuk:2022msa}
O.~Savchuk, R.~V.~Poberezhnyuk, A.~Motornenko, J.~Steinheimer, M.~I.~Gorenstein and V.~Vovchenko,
Phys. Rev. C \textbf{107}, no.2, 024913 (2023)

\bibitem{Savchuk:2022ljy}
O.~Savchuk, R.~V.~Poberezhnyuk and M.~I.~Gorenstein,
Phys. Lett. B \textbf{835}, 137540 (2022)

\bibitem{Begun:2004pk}
V.~V.~Begun, M.~I.~Gorenstein, A.~P.~Kostyuk and O.~S.~Zozulya,
Phys. Rev. C \textbf{71}, 054904 (2005)

\bibitem{Bzdak:2012ab}
A.~Bzdak and V.~Koch,
Phys. Rev. C \textbf{86}, 044904 (2012)

\bibitem{Savchuk:2019xfg}
O.~Savchuk, R.~V.~Poberezhnyuk, V.~Vovchenko and M.~I.~Gorenstein,
Phys. Rev. C \textbf{101}, no.2, 024917 (2020)

\bibitem{Kuznietsov:2022pcn}
V.~A.~Kuznietsov, O.~Savchuk, M.~I.~Gorenstein, V.~Koch and V.~Vovchenko,
Phys. Rev. C \textbf{105}, no.4, 044903 (2022)

\bibitem{Vovchenko:2020gne}
V.~Vovchenko, R.~V.~Poberezhnyuk and V.~Koch,
JHEP \textbf{10}, 089 (2020)

\bibitem{Barej:2022jij}
M.~Barej and A.~Bzdak,
Phys. Rev. C \textbf{106}, no.2, 024904 (2022)

\bibitem{Vovchenko:2021yen}
V.~Vovchenko,
Phys. Rev. C \textbf{105}, no.1, 014903 (2022)

\bibitem{Barej:2022ccb}
M.~Barej and A.~Bzdak,
Phys. Rev. C \textbf{107}, no.3, 034914 (2023)

\bibitem{Poberezhnyuk:2020cen}
R.~V.~Poberezhnyuk, O.~Savchuk, M.~I.~Gorenstein, V.~Vovchenko and H.~Stoecker,
Phys. Rev. C \textbf{103}, no.2, 024912 (2021)


\bibitem{OmanaKuttan:2022the}
M.~Omana Kuttan, A.~Motornenko, J.~Steinheimer, H.~Stoecker, Y.~Nara and M.~Bleicher,
Eur. Phys. J. C \textbf{82}, no.5, 427 (2022)

\bibitem{Steinheimer:2010ib}
J.~Steinheimer, S.~Schramm and H.~Stocker,
J. Phys. G \textbf{38}, 035001 (2011)

\bibitem{Steinheimer:2011ea}
J.~Steinheimer, S.~Schramm and H.~Stocker,
Phys. Rev. C \textbf{84}, 045208 (2011)

\bibitem{Motornenko:2019arp}
A.~Motornenko, J.~Steinheimer, V.~Vovchenko, S.~Schramm and H.~Stoecker,
Phys. Rev. C \textbf{101}, no.3, 034904 (2020)

\bibitem{HADES:2020wpc}
J.~Adamczewski-Musch \textit{et al.} [HADES],
Phys. Rev. C \textbf{102}, no.2, 024914 (2020)



\end{thebibliography}
\end{document}